\documentclass{amsart}

\usepackage{graphicx}
\usepackage{color}
\newtheorem{theorem}{Theorem}%[section]
 \newtheorem{lem}{Lemma}%[section]
 \newtheorem{prop}{Proposition}%[section]
 \newtheorem{defn}{Definition}%[section]
  \newtheorem{remark}{Remark}%[section]
  \newtheorem{example}{Example}%[section]
\newtheorem{corollary}{Corollary}
\UseRawInputEncoding
\usepackage[russian, english]{babel}
\def\bd{{\bf d}}
\def\br{{\bf r}}
\def\dfrac#1#2{\displaystyle{#1\over #2}}

\def\bv{{\bf V}}
\def\bV{{\bf V}}

\def\Div{\mbox{div}\,}

\def\bB{{\bf B}}

\def\bE{{\bf E}}
\begin{document}

%\markboth{ Rozanova}{Formation of singularities of the solution in a two-dimensional problem
%of oscillations of an electron plasma}
% in the Model of Upper Hybrid Oscillations}

%%%%%%%%%%%%%%%%%%% Publisher's Area please ignore %%%%%%%%%%%%%%%%%%%%%%%
%
%\catchline{}{}{}{}{}
%
%%%%%%%%%%%%%%%%%%%%%%%%%%%%%%%%%%%%%%%%%%%%%%%%%%%%%%%%%%%%%%%%%%%%%%%%%%

\title[ %two-dimensional
 Symmetric solutions of the Euler-Poisson equations]{
On the behavior of multidimensional radially symmetric solutions of
the repulsive Euler-Poisson equations
%Interaction of Regularizing Factors in the Model of Upper Hybrid Oscillations of Cold Plasma
}

\author{Olga S. Rozanova}

\address{ Mathematics and Mechanics Department, Lomonosov Moscow State University, Leninskie Gory,
Moscow, 119991,
Russian Federation,
rozanova@mech.math.msu.su}



\maketitle

%\begin{history}
%\received{(Day Month Year)}
%\revised{(Day Month Year)}
%\accepted{(Day Month Year)}
%\comby{(xxxxxxxxxx)}
%\end{history}

\begin{abstract}
It is proved that the radially symmetric solutions of the repulsive
Euler-Poisson equations with a non-zero background, corresponding to
cold plasma oscillations blow up in many spatial dimensions except for $\bd=4$ for
almost all initial data. The initial data, for which the solution may not blow up,
correspond to simple waves. Moreover, if a solution is globally smooth
in time, then it is either affine or tends to affine as
$t\to\infty$.
 % except for
%the affine ones. In particular, arbitrary small perturbations of the
%affine solution blows up.
\end{abstract}

\section{Introduction}

In this paper, we study a version of the repulsive Euler-Poisson
equations
\begin{eqnarray}\label{EP}
\dfrac{\partial n }{\partial t} + \Div(n \bv)=0,\quad
\dfrac{\partial \bv }{\partial t} + \left( \bv \cdot \nabla \right)
\bv =\,k \,  \nabla \Phi, \quad \Delta \Phi =n-n_0,
\end{eqnarray}
where the solution components, the scalar functions $n$ (density),
$\Phi$ (force potential), and the vector $\bv$ (velocity) depend on
the time $t$ and the point $x\in {\mathbb R}^\bd $, $\bd\ge 1$,
$n_0> 0$ is the density background. Positive or negative value of
the  constant $k$ corresponds to the repulsive and attractive force,
respectively.

The Euler-Poisson equations arise in many applications, see
\cite{ELT} for references, but for us they are of interest primarily
in the context of cold plasma oscillations, where  $k =1$. At
present, much attention is paid to the study of cold plasma in
connection with the possibility of accelerating electrons in the
wake wave of a powerful laser pulse \cite {esarey09}; nevertheless,
there are very few theoretical results in this area.

The equations of hydrodynamics of ``cold" or electron plasma in the
non-relativistic approximation in dimensionless quantities have the
form (see, e.g.,~\cite{ABR78},  \cite {GR75})
\begin{eqnarray}\label{base1.1}
\dfrac{\partial n }{\partial t} + \Div(n \bv)=0,\quad
\dfrac{\partial \bv }{\partial t} + \left( \bv \cdot \nabla \right) \bv
= \, - \bE -  \left[\bv \times  \bB\right],\\
\frac{\partial \bE }{\partial t} =   n \bv
 + {\rm rot}\, \bB,\qquad %\label{base1.3}\\
\frac{\partial \bB }{\partial t}  =
 - {\rm rot}\, \bE,\qquad \Div \bB=0,\label{base1.5}
\end{eqnarray}
$ n$ and $ \bv=(V_1, V_2, V_3)$  are the density and velocity of electrons,
$ \bE=(E_1, E_2, E_3)$ and  $  \bB=(B_1, B_2, B_3) $  are vectors of electric and magnetic fields.
All components of solution depends on $t\in {\mathbb R}_+$ and $x\in{\mathbb R}^3$.

It is commonly known that the plasma oscillations described by
\eqref{base1.1}, \eqref{base1.5}, tend to break. Mathematically,
 the breaking process means a blow-up  of the solution, and the appearance
 of a delta-shape singularity of the electron density \cite
 {david72},
see also   \cite{CH18} and references therein for numerical
 examples illustrating the behavior of solution on the stage of the
 blow-up.
 Among the main interests of physicists is the study of the possibility of the existence
 of a smooth solution for as long as possible.
%\bigskip

%\section{Equations for axially symmetric oscillations}
System \eqref{base1.1} -- \eqref{base1.5} has an important
class of solutions depending only on the radius-vector of point
$r=\sqrt{x_1^2+x_2^2+x_3^2}$, i.e.
\begin{eqnarray}\label{sol_form}
\bv=F(t,r) \br,\quad \bE=G(t,r) \br,\quad \bB=Q(t,r)\br, \quad
n=n(t,r),
\end{eqnarray}
where $ \br=(x_1, x_2,x_3)$.

Since
 ${\rm div} \bB = 3 Q(t,r)+ r(Q(t,r))_r,$ then the condition ${\rm div} \bB =
 0$ implies $Q(t,r)=\tilde Q(t) r^{-3}$, therefore a bounded in the
 origin solution exists if and only if $ \bB \equiv 0. $ In its turn, it implies ${\rm rot}\,
 \bE=0$.
 From the first equations of \eqref{base1.1} and  \eqref{base1.5} under the assumption
 that the solution is sufficiently smooth and that the steady-state density $n_0$ is equal to 1, we
 get
\begin{eqnarray}
n=1- \Div \bE,\label{n}\end{eqnarray} therefore $ n $ can be removed
from the system. Thus, the resulting system is
\begin{eqnarray}\label{4}
\dfrac{\partial \bv }{\partial t} + \left( \bv \cdot \nabla \right)
\bv = \, - \bE,\quad \frac{\partial \bE }{\partial t} + \bv \Div \bE
 = \bV.%\quad{\rm rot}\, \bE=0
\end{eqnarray}

If we introduce the potential $\Phi$ such that $\nabla \Phi = -\bE$,
we can rewrite system \eqref{n}, \eqref{4} as the Euler-Poisson
equations \eqref{EP} with $n_0=1$.

Note that system \eqref{4} can be considered in any space dimensions
$\bd$ (non necessarily $\bd=3$, as in the initial setting). In what
follows, we deal with just this case.

 Consider the initial data
\begin{equation}\label{CD1}
(\bv, \bE) |_{t=0}= (F_0(r) {\bf r}, G_0(r) {\bf r} ), \quad (F_0(r)
, G_0(r) ) \in C^2(\bar {\mathbb R}_+),
\end{equation}
where $ {\bf r}=(x_1, \dots ,x_\bd), \quad r=|{\bf r}|,$ with the
physically natural condition $n|_{t=0}>0$.

We call a solution of \eqref{4},  \eqref{CD1} smooth for $t\in
[0,t_*)$, $t_*\le\infty$, if the functions $F$ and $G$ in
\eqref{sol_form} belong to the class $ C^1([0,t_*)\times
\bar{\mathbb R}_+)$. The blow-up of solution implies that the
derivatives of solution tends to infinity as $t\to t_*<\infty$.

\begin{defn}
Solution $(\bv, \bE)$ to system \eqref{4} is called an affine
solution if it has the form $\bv=\mathfrak{V}(t){\bf r}$,
$\bE=\mathfrak{E}(t){\bf r}$, where $\mathfrak{V}$ and
$\mathfrak{E}$ are $(\bd\times \bd)$ matrices.
\end{defn}

\begin{defn} Solution $(\bv, \bE)$ to system \eqref{4} is called a simple wave
if it has the form $\bv=F(t,r){\bf r}$,
$\bE=G(t,r){\bf r}$, where $F(t,r)$ and
$G(t,r)$ are functionally dependent.
\end{defn}

The main result of this paper is the following:
 \begin{theorem}\label{MT} The solution of the Cauchy problem
\eqref{4}, \eqref{CD1} for $\bd\ge 2$, $\bd \ne 4$, blows up in a finite time for
all initial data, possibly except for the data, corresponding to simple waves. If the solution is globally smooth in
time, then it is either affine or tends in the $C^1$-norm on each
compact subset of the half-axis $[0,\infty)$ to an affine solution
as $t\to \infty$.
\end{theorem}

Theorem \ref{MT} can be reformulated in the terms of the Euler-Poisson
equations \eqref{EP}.

The results are obtained analytically for small deviations of the trivial steady state and partly thanks
to numerical considerations for arbitrary initial data.

\medskip

The meaning of Theorem \ref{MT}  is that in spaces of higher dimensions (except for $\bd=4$)
there cannot be a non-trivial solution of the Euler-Poisson equations, globally smooth in time, for practically all initial data.
In other words, if we want to get a globally smooth solution, we must choose the initial data in a very narrow class of"simple waves,
for which one  component the solution depends on another. Such a class of solutions cannot be realized as a perturbation of a stationary state with a compact support (see  Sec.\ref{SV}), which makes them unpromising both from the point of view of physical applications and from the point of view of a computational experiment.

Note that many results can be obtained in a more convenient way for
the Euler-Poisson equations written as \eqref{4}, this applies, in
particular, to equations with linear damping, cf. \cite{ChRD2020}
and \cite{BL}.

An important contribution to the study of critical phenomena of
solutions of the repulsive Euler-Poisson equations was made in
\cite{ELT}, where, in particular, radially symmetric solutions in a
space of many dimensions are studied. In \cite{WTB} a critical
threshold in the terms of the initial data was obtained for this
case. However, the background value of the density was taken as zero
in all these studies. In our case, a nonzero background density
naturally arises, which leads to the appearance of oscillations and
radically changes the research technique and the behavior of the
solution. Let us note \cite{T21}, which contains some progress in this case.

Note that the behavior of the solution changes significantly in the
case attractive case $k<0$ , we refer to \cite{CT}, where conditions
for the existence of a globally smooth solution in terms of initial
data for zero and non-zero backgrounds are obtained. See also
\cite{BL} for recent interesting results concerning critical
thresholds in Euler-Poisson systems in different settings.

\section{Explicit solutions along characteristics}\label{S2}
 First of all, we notice that  \eqref{sol_form} and \eqref{4}
imply that $F$ and $G$ satisfy the following Cauchy problem:
 \begin{eqnarray}\label{sys_pol1}
    \dfrac{\partial G}{\partial t}+F r \dfrac{\partial G}{\partial r}=F-\bd F
    G, \quad
    \dfrac{\partial F }{\partial t}+F r \dfrac{\partial F}{\partial r}=-F^2 -
    G,
 \end{eqnarray}
\begin{equation}\label{CD2}
(F(0,r), G(0,r))=(F_0(r), G_0(r)), \quad (F_0(r) , G_0(r) ) \in
C^2(\bar {\mathbb R}_+).
\end{equation}

 Along the characteristic
 \begin{eqnarray}\label{char}
 \dot r = F r,
 \end{eqnarray}
 starting from the point $r_0\in [0,\infty)$
 system \eqref{sys_pol1} takes the form
  \begin{eqnarray}\label{GF}
\dot G =F-\bd F
    G,\qquad \dot{F } =  -F^2 -
    G.
\end{eqnarray}
 On the phase plane $(G,F)$ system \eqref{GF} implies one equation
\begin{eqnarray*}
\frac12 \dfrac{d F^2}{d G}=-\frac{ F^2 +
    G}{1-\bd
    G},
\end{eqnarray*}
which is linear with respect to $F^2$ and can be explicitly integrated. Indeed,
we have for $\bd=2$
\begin{eqnarray}\label{fiRad2}
2 F^2=(2G-1) \ln|1-2G|+C_2(2G-1)  -1, \\ C_2=\frac{1+2
F^2(0,r_0)}{2G(0,r_0)-1}-\ln|1-2G(0,r_0)|,\nonumber
\end{eqnarray}
for $\bd=1$ and $\bd\ge 3$
\begin{eqnarray}\label{fiRad3}
&&F^2=\frac{2 G-1}{\bd-2}+C_\bd |1-\bd G|^\frac{2}{\bd} , \quad
C_\bd=\frac{1-2 G(0,r_0)+(\bd-2) F^2(0,r_0)}{(\bd-2)|1-\bd
G(0,r_0)|^\frac{2}{\bd}}.
\end{eqnarray}

\bigskip
\begin{lem}\label{P1}
 The condition
$n>0$ implies $G< \frac{1}{\bd}$ provided $G(t,r)\in
C^1([0,t_*)\times \bar{\mathbb R})$, $t_*\le \infty$.
\end{lem}

\proof From the first equation \eqref{GF} and \eqref{char} we have
\begin{equation}\label{rG}
1-\bd\,G={\rm const}\, r^{-\bd},
\end{equation}
therefore the sign of $1-\bd \, G(r(t))$ coincides with the sign of
$1-\bd G(0,r_0)$, this means that on the phase plane $(G,F)$ of the
system \eqref{GF} the motions on the half-planes $G<\frac{1}{\bd}$
and $G>\frac{1}{\bd}$ are separated.

Since \eqref{n} implies $\Div \bE=\bd\,G+G_r r<1$, then at the point
$r=0$ the condition $G< \frac{1}{\bd}$ follows from \eqref{n}
directly. Further, on the phase plane $(G,F)$  there is only one
equilibrium, the origin. Therefore, in the half-plane
$G>\frac{1}{\bd}$ no bounded trajectory exists. If in some point
$r_0>0$ initially $\bd\, G(0,r_0)>1$, then $G(r(t))\to +\infty$ and
$r(t)\to 0$ as $t\to t_*>0$ (see \eqref{rG}), and we obtain a
contradiction with  property $n>0$.  $\Box$

\bigskip
\begin{lem}\label{P2}
For $\bd\ge 2$ the functions $F$ and $G$ are bounded if and only if
$G_0(r)<\frac{1}{\bd}$.
\end{lem}

\proof If $G(0,r_0)<\frac{1}{\bd}$, $r_0\in \bar {\mathbb R}$, then
the phase trajectories of \eqref{GF} are in the half-plane
$G<\frac{1}{\bd}$ (Lemma \ref{P1}). In this half-plane the leading
term in the right hand side of \eqref{fiRad2} is $(2\,G-1)
\ln|1-2\,G|$, and the leading term in the right hand side of
\eqref{fiRad3} is $\frac{2\, G-1}{\bd-2}$.  Therefore $F$ and $G$
are bounded for any $t>0$. $\Box$

\bigskip Note that the situation is different for $\bd=1$, where the
phase trajectory is bounded not for all $G<\frac{1}{\bd}=1$. Indeed,
the leading term as $G\to-\infty$ in the right hand side of
\eqref{fiRad3} is $C_1 (1-G)^2$, $F$ and $G$ can be bounded if only
if $\displaystyle C_1=\frac{F^2(0)+2G(0)-1}{(1-G(0))^2}<0$ for every $r_0$. The 1D
case is completely analyzed in \cite{RChZAMP21}.

\bigskip

Our considerations also imply
\begin{lem}\label{P3}
 Let $\bd\ge 2$. Then along every characteristic curve starting from $r_0\in\bar{\mathbb R}_+$ the functions
 $F(t)$ and $G(t)$, solution of \eqref{GF}, are periodic with the period
\begin{equation}\label{T}
  T= 2 \int\limits_{G_-}^{G_+} \frac {d\eta}{(1-\bd\, \eta )F(\eta)},
\end{equation}
$F$ is given as \eqref{fiRad2} or  \eqref{fiRad3}, $G_-<0$ and
$G_+>0$ are the lesser and greater roots of the equation $F(G)=0$.
Moreover, $\int\limits_0^{T} F(\tau) \, d\tau=0$.
\end{lem}

The last property follows from the fact that  the phase curve of
\eqref{GF} is symmetric with respect to the axis $F=0$.

The following lemma helps us to study the properties of the period $T$ depending on $\bd$.
\begin{lem}\label{L4} The period of revolution of the phase curve of  equation \eqref{GF}  depends on $\bd$ and the starting point of trajectory, except for $\bd=1$ and $\bd=4$, where $T=2\pi$.  In the other cases the following asymptotics holds for the deviation of order $\varepsilon$ from the origin:
\begin{equation}\label{T_per}
T=2\pi (1+ \frac{1}{24} (\bd-1)(\bd-4) \varepsilon^2+o(\varepsilon^2)),\quad \varepsilon\to 0,
\end{equation}
 i.e. for $\bd \in (1, 4)$ the period is less that $2\pi$, for $\bd > 4$ the period is greater that $2\pi$.
\end{lem}

\proof The results for $\bd=1$ and $\bd=4$ can be obtained explicitly.  For $\bd=1$ the period was computed in \cite{RChZAMP21}, for $\bd=4$
$$I_4=\int \limits^G \frac{d\eta}{(1-\bd\, \eta )F(\eta)}=
\arctan\frac{1-2 C_4 \sqrt{1-4 G} }{\sqrt{4 C_4 \sqrt{1-4G} +4G-2}} +\rm const.$$
For a starting point $(G_+,0))$, $G_+\in \left(0, \frac{1}{\bd}\right)$, we get
$C_4=\frac{1-2 G_+}{2\sqrt{1-4 G_+}}
$ and
$G_-=\frac{ G_+}{4 G_+-1}<0$.  Since $\lim\limits_{G\to G_+-0} I_4  - \lim\limits_{G\to G_-+0} I_4 =\pi$,  the period is $2 \pi$.

To prove formula \eqref{T_per} we use the Lindstedt-Poincar\'e method of stretching of the independent coordinate to avoid secular terms in regular asymptotic expansions (e.g. \cite{Naife}, Sec.3). Namely, we choose a small parameter $\varepsilon$, the deviation of the initial point of the trajectory from the origin,  set
\begin{eqnarray}\label{ts}
t=s \, (1+\varepsilon w_1+\varepsilon^2 w_2+o(\varepsilon^2))
\end{eqnarray}
together with
\begin{eqnarray*}
G(s)= \varepsilon G_1(s)+\varepsilon^2 G_2(s)+ \varepsilon^3 G_3(s)+ o(\varepsilon^3), \quad
F(s)= \varepsilon F_1(s)+\varepsilon^2 F_2(s)+ \varepsilon^3 F_3(s)+ o(\varepsilon^3),
\end{eqnarray*}
and substitute to \eqref{GF}. Thus, for initial point $(\varepsilon, 0)$ we get
\begin{eqnarray*}
G_1(s)& =& \cos s,\quad F_1(s) = -\sin s, \\
G_2(s)&=&- w_1 \, s \sin s  -\frac12+\frac13(2+ \bd) \cos s- \frac13(2+ \bd )\cos 2 s,\\
F_2(s)&=& - w_1 \, s \cos s -\frac13(2+ \bd) \sin s+\frac13(2+ \bd)  \sin 2 s,
\end{eqnarray*}
what implies $w_1=0$, and
\begin{eqnarray*}
G_3(s)&=& a \,s \sin s +A_0+A_1 \cos s + A_2 \cos 2 s + A_3 \cos 3 s,\\
F_3(s)&=& a\, s \cos s +B_1 \sin s+ B_2 \sin 2 s+ B_3 \sin 3 s,
\end{eqnarray*}
where $a=\frac{1}{24} (\bd-1)(\bd-4)-w_2 $ and $A_i=A_i$, $i=0,\dots, 3$,  $B_i=B_i$, $i=1,\dots, 3$, are constant that depend on $\bd$. Thus, to eliminate the secular term, we must choose $
a=0$, or $w_2=\frac{1}{24} (\bd-1)(\bd-4)$. Thus, returning to the variable $t$, we obtain  formula
\eqref{T_per}.
$\Box$

\begin{remark}
Continuing the calculation of subsequent coefficients in the expansion \eqref{ts} at each step we get functions $G_i(s)$ and $F_i(s)$ having  period $2 \pi$ with respect to $s$, and $w_i= (\bd-1)(\bd-4) W_i(\bd)$, where $W_i(\bd)$ is a function, positive for $\bd>0$.
\end{remark}

Fig.1 shows the phase curves on the  plane $(F,G)$ for the same starting point $(0.1, 0)$ and the period $T$ for different starting points $(G_+, 0)$ (found numerically as improper integral \eqref{T})
for different $\bd$. Evidently,
 $T(0)=2\pi$ (see \eqref {P} with $G=F=0$) and $T\to \pi  \sqrt{ \bd}=2\int\limits_0^\infty \frac{dF}{F^2+\frac{1}{\bd}}$
 as $G_+\to \frac{1}{\bd}$;

 %, the numerical result is presented in Fig.2 (left).

\begin{figure}[htb!]
\hspace{-0.5cm}
\begin{minipage}{0.4\columnwidth}
%\centerline{
\includegraphics[scale=0.7]{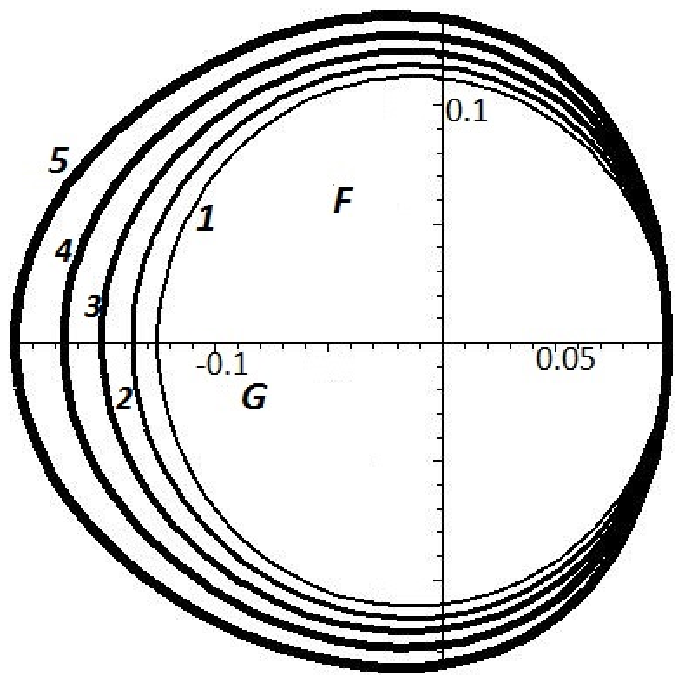}
%\vspace{-0.5 cm}
\end{minipage}
\hspace{2.0cm}
\begin{minipage}{0.4\columnwidth} %\centerline{
\includegraphics[scale=0.7]{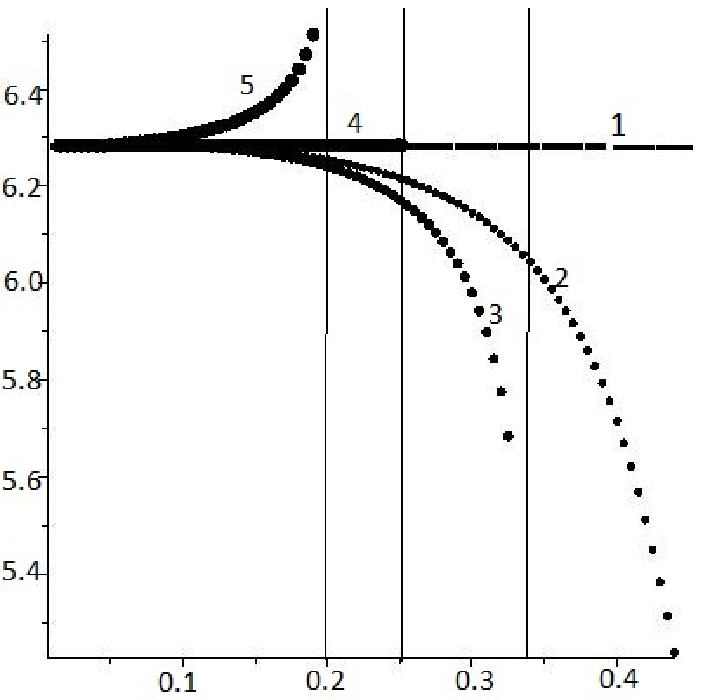}
%\vspace{-0.5 cm}
\end{minipage}
\caption{Phase portrait on the  plane $(F,G)$ for $G_+=0.1$ (left) and  the dependence of the period on $G_+$ (right) for $\bd =1, 2, 3, 4, 5$.  }\label{L2}
\end{figure}

\section{The behavior of derivatives}\label{Der}

 We denote ${\mathcal D}=\Div \bv$,  $\lambda=\Div \bE$,
$J_{ij}=\partial_{x_i}{V_i}\,\partial_{x_j}{V_j}-\partial_{x_j}{V_i}\,\partial_{x_i}{V_j}
%\det(\|\partial_{x_i}{V_j}\|),$ $i,j=1,\dots, \bd
$,
$i\ne
j$,
 $J=\sum\limits_{i,j=1,\, i\ne j}^\bd J_{ij}.$ The number of terms in the
 sum is $C^\bd_2=\frac{(\bd-1)\bd}{2}$.

It can be readily shown that
\begin{eqnarray}\label {DJ}
 &&{\mathcal D} = \bd \, F +F_r \, r, \quad  \lambda = \bd\, G +G_r \,
 r,%\\\nonumber
 \quad J=(\bd-1)\,F \,F_r\,r+\frac{(\bd-1)\bd}{2}\, F^2,
\end{eqnarray}
therefore
\begin{eqnarray}\label{2J}
  2J &=&2\,(\bd-1)\,{\mathcal D}\, F-(\bd-1)\,\bd \,F^2.
\end{eqnarray}

\bigskip

\begin{lem}\label{P4} %Assume $d\ge 2$ and $G_0(r)\le \frac{1}{d}$.
The solution of the Cauchy problem \eqref{sys_pol1}, \eqref{CD2}
with a bounded density belongs to the class  $C^1([0, t_*)\times
\bar{\mathbb R}_+)$ if and only if  $\mathcal D$ and $\lambda$ are
bounded on $[0, t_*)\times \bar{\mathbb R}_+$, $t_*\le \infty$.
\end{lem}
\proof If the density is bounded, then, as follows from Lemma \ref{P2},
$F$ and $G$ are bounded. If the solution belongs to the class  $C^1([0, t_*)\times
\bar{\mathbb R}_+)$, then $F_r$ and $G_r$ are bounded for $t\in [0,t_*)$.
In turn,  \eqref{DJ} implies that $\mathcal D$ and $\lambda$ are bounded.

On the other hand, if
$\mathcal D$ and $\lambda$ are bounded together with the density, then  $F$ and $G$ are bounded by Lemma \ref{P2}  and
$F_r$ and $G_r$ are bounded from \eqref{DJ}.
Since \eqref{sys_pol1} is
symmetric hyperbolic system, this implies that the solution is $C^1$-
smooth globally in $t$  (see \cite{Daf16}). If $\mathcal D$ or $\lambda$ tends to
$\infty$ as $t \to t_*<\infty$, then a singularity forms at the
point $t_*<\infty$. $\Box$

\bigskip

 System \eqref{4} implies
\begin{eqnarray*}\label{Dlam}
\dfrac{\partial {\mathcal D}}{\partial t} &+& ( \bv \cdot \nabla
{\mathcal D} ) =   - {\mathcal D}^2 + 2J -\lambda,\qquad
\dfrac{\partial \lambda }{\partial t} + ( \bv \cdot \nabla  \lambda
) ={\mathcal D}(1-\lambda).
\end{eqnarray*}

Taking into account \eqref{2J},
 along the characteristic curve,
starting from the point $r_0$, we obtain
\begin{eqnarray}\label{lamD}
 \dot{\mathcal D } =   - {\mathcal D}^2 + 2 \,(\bd-1)\,F\, {\mathcal D}-\lambda -(\bd-1)\,\bd \, F^2,\qquad
\dot \lambda ={\mathcal D}\,(1-\lambda),
\end{eqnarray}
a quadratically nonlinear system with the coefficient $F$ found from
\eqref{GF}. In fact, \eqref{GF}, \eqref{lamD}, can be considered as
a system of 4 ODEs for $G, F, {\mathcal D}, \lambda$, where
\eqref{GF} is separated.

\bigskip
Let us introduce new variables: $u={\mathcal D}-\bd\, F$,
$v=\lambda-\bd\, G$. Evidently, if $u, v, F, G$ are bounded, then
${\mathcal D}$ and $\lambda$ are bounded. Thus, we get a system
\begin{equation}\label{uv}
    \dot u=-u^2-2\,F\, u  -v, \quad \dot v=-u\,v + (1-\bd\, G) \,u- \bd\, F\,
    v.%\quad \dot
    %{\phantom{u}}= \dfrac{\partial }{\partial t} + r F \dfrac{\partial }{\partial
    %r}.
\end{equation}

\begin{lem} Assume $\bd \ge 2$.
If the data \eqref{CD1} are such that the solution to the Cauchy
problem \eqref{uv},
$$u_0(r_0)=({\mathcal D}-\bd F)|_{t=0,\,
r=r_0},\quad  v_0(r_0)=({\lambda}-\bd G)|_{t=0,\, r=r_0},  $$ is
bounded for all fixed $r_0\in [0,+\infty)$, $t\in [0, t_*)$, $t_*\le
\infty$, then ${\mathcal D}$ and $\lambda$ are bounded and for $t\in
[0, t_*)$  there exists a $C^1$-smooth  solution to \eqref{4},
\eqref{CD1} with a positive density.
\end{lem}

\proof The lemma is a corollary of Lemmas \ref{P1}, \ref{P2} and
\ref{P4}. Indeed, if the solution \eqref{uv} is bounded, then ${\mathcal D}$ and $\lambda$ are bounded,
since Lemmas \ref{P1} and \ref{P2} imply boundedness of $G$ and $F$ and positivity of density. According to the lemma \ref{P4}, there exists a $C^1$-smooth solution to the problem \eqref{4}, \eqref{CD1} with positive density.
 $\Box$

\begin{corollary}
In the case $\bd\ge 2$ the affine solution to \eqref{4}, \eqref{CD1}
is a globally smooth solution with  positive density.
\end{corollary}

\proof System \eqref{uv} has the trivial solution, it is evidently
globally smooth. It corresponds to the affine solution, since $u=F_r
r=0$ and $v=G_r r=0$  imply $F=f(t)$ and $G=g(t)$. The affine solution has the form
$\bv=\mathfrak{V}(t){\bf r}$,
$\bE=\mathfrak{E}(t){\bf r}$, with $\mathfrak{V}(t)= f(t) \mathfrak I$, $\mathfrak{E}(t)= g(t) \mathfrak I$, $\mathfrak I$ is the identity matrix. $\Box $.

\begin{remark}
The fact that the affine solution with  positive density is globally
smooth for $\bd\ge 2$, follows directly from \eqref{fiRad2},
\eqref{fiRad3}. Indeed, if $F(t,r)=f(t)$, $G(t,r)=g(t)$, then $F$
and $G$ are derivatives of the solution $(\bV= F\,\br, \bE=G\,\br)$ with respect to $|\br|$, and we proved  in Sec.\ref{S2} that
they are bounded, if $g(0)<\frac{1}{\bd}$, i.e. density is positive.
\end{remark}

\bigskip

\section{Linearization and the Radon lemma}\label{SRadon}

Next, we need the following version of the Radon lemma (1927)
\cite{Riccati}, Theorem 3.1, see also \cite{Radon}.

\begin{theorem}[The Radon lemma]
\label{T2} A matrix Riccati equation
\begin{equation}
\label{Ric}
 \dot W =M_{21}(t) +M_{22}(t)  W - W M_{11}(t) - W M_{12}(t) W,
\end{equation}
 {\rm (}$W=W(t)$ is a matrix $(n\times m)$, $M_{21}$ is a matrix $(n\times m)$, $M_{22}$ is a matrix  $(m\times m)$, $M_{11}$ is a matrix  $(n\times n)$, $M_{12} $ is a matrix $(m\times n)${\rm )} is equivalent to the homogeneous linear matrix equation
\begin{equation}
\label{Lin}
 \dot Y =M(t) Y, \quad M=\left(\begin{array}{cc}M_{11}
 & M_{12}\\ M_{21}
 & M_{22}
  \end{array}\right),
\end{equation}
 {\rm (}$Y=Y(t)$  is a matrix $(n\times (n+m))$, $M$ is a matrix $((n+m)\times (n+m))$ {\rm )} in the following sense.

Let on some interval ${\mathcal J} \in \mathbb R$ the
matrix-function $\,Y(t)=\left(\begin{array}{c}\mathfrak{Q}(t)\\ \mathfrak{P}(t)
  \end{array}\right)$ {\rm (}$\mathfrak{Q}$  is a matrix $(n\times n)$, $\mathfrak{P}$  is a matrix $(n\times m)${\rm ) } be a solution of \eqref{Lin}
  with the initial data
  \begin{equation*}\label{LinID}
  Y(0)=\left(\begin{array}{c}I\\ W_0
  \end{array}\right)
  \end{equation*}
   {\rm (}$ I $ is the identity matrix $(n\times n)$, $W_0$ is a constant matrix $(n\times m)${\rm ) } and  $\det \mathfrak{Q}\ne 0$ on ${\mathcal J}$.
  Then
{\bf $ W(t)=\mathfrak{P}(t) \mathfrak{Q}^{-1}(t)$} is the solution of \eqref{Ric} with
$W(0)=W_0$ on ${\mathcal J}$.
\end{theorem}

System \eqref{uv} can be written as \eqref{Ric} with
%$$
\begin{eqnarray*}\label{M}
W=\begin{pmatrix}
  u\\
  v
\end{pmatrix},\quad
M_{11}=\begin{pmatrix}
  0\\
\end{pmatrix},\quad
 M_{12}=\begin{pmatrix}
  1 & 0\\
\end{pmatrix},\\
M_{21}=\begin{pmatrix}
 0 \\ 0
\end{pmatrix},\quad
M_{22}=\begin{pmatrix}
  - 2\,F& -1\\
  1-d\,G  & -d\,F\\
\end{pmatrix}.\\\nonumber
%$$$$
\end{eqnarray*}

Thus, we obtain the linear Cauchy problem
\begin{eqnarray}
\label{matr}
 \begin{pmatrix}
  \dot q\\
  \dot p_1\\
  \dot p_2\\
\end{pmatrix}
=\begin{pmatrix}
0& 1& 0\\
0&- 2\,F& -1\\
0& 1-d\,G  & -d\,F\\
\end{pmatrix}
\begin{pmatrix}
  q\\
  p_1\\
  p_2\\
\end{pmatrix},\quad
\begin{pmatrix}
  q\\
  p_1\\
  p_2
  \\
\end{pmatrix}(0)=\begin{pmatrix}
  1\\
  u_0\\
  v_0\\
\end{pmatrix},
\end{eqnarray}
with periodical coefficients, known from \eqref{GF}. System
\eqref{matr} implies %the linear ODE with respect to to $p_1$:
\begin{eqnarray*}\label{p1}
  \ddot p_1+(2+\bd) F \dot p_1 +(1+2(\bd-1)F^2 -(2+\bd) G) p_1=0.
\end{eqnarray*}
The standard change of the variable $p_1(t)=P(t)\, e^{-
\frac{\bd+2}{2}\int\limits^t_0 F(\tau)\,d\tau}$ reduces the latter
equation to% \eqref{p1}
%the linear second order ODE
\begin{eqnarray}\label{P}
  \ddot P+{Q} P=0, \quad {Q}=1-\frac{\bd+2}{2} G-\frac14 (\bd-2)(\bd-4)F^2.
\end{eqnarray}

Theorem \ref{T2} implies that the solution of \eqref{uv} blows up if
and only if $q(t)$ vanishes at some point $t_*,\,0<t_*<\infty$.

From \eqref{matr} we find
\begin{equation}\label{q}
 q(t)=1+\int\limits_0^t p_1(\tau)\,d
\tau=1+\int\limits_0^t P(\xi)\, e^{- \frac{d+2}{2}\int\limits^\xi_0
F(\tau)\,d\tau}\,d\xi.
\end{equation}

\section{Proof of the main theorem}

1. The idea of the proof of the theorem is as follows. It is shown that in the case of initial data of a general form, not related to a simple wave, the function $q$ contains a term that is the product of a periodic function and a growing exponent, and therefore necessarily vanishes in a finite time. Since $q$ serves as the denominator in the expression for $u$ and $v$, the derivative of the solution $(\bV, \bE)$ becomes unbounded in finite time.
Since $q$ is expressed in terms of $P$ and $F$ (see \eqref{q}), and $F$ is periodic with zero-mean periodicity (see Lemma 3), the property of $q$ to oscillate with increasing amplitude is inherited on the properties of solutions to equation \eqref{P}, which is an ordinary differential equation with periodic coefficients.

Such equations are described by the Floquet theory (e.g.\cite{Chicone}).
It implies that for a periodic  ${Q}(t)$ with period $T$, any solution of  \eqref{P} has
the form $P={\mathcal P}_1(t) e^{\sigma_1 t} + {\mathcal P}_2(t)
e^{\sigma_2 t}$, $\sigma_1\ne \sigma_2$, or $P=e^{\sigma t} ({\mathcal P}_1 (t)+ {\mathcal
P}_2(t)\, t)$, where the functions ${\mathcal P}_1, {\mathcal P}_2$
are periodic with period $T$.
However, in the general case there is no methods of finding characteristic exponents.  % $\sigma_i$.
  In our case ${Q}(t)={Q}(-t)$, therefore \eqref{P} has solutions $e^{\mu t} \mathcal P(t)$, and $e^{-\mu t} \mathcal P(-t)$, $\mathcal P$
 is $T$-periodic, which can be taken as a fundamental system provided that $\mu$ is real.

2. Let us describe the idea of looking for $\mu$ (\cite{BEr67}, section 16.2). Suppose $z(t)$ is a solution of \eqref{P} with
initial conditions
 $z(0)=1$, $z'(0)=0$. Then $$z(t)=\frac{1}{2\mathcal P(0)}\left(
e^{\mu t} \mathcal P(t)+e^{-\mu t} \mathcal P(-t)
 \right), $$
 since $\mathcal P(\pm T)=\mathcal P(0)$, $\mathcal P'(\pm T)=\mathcal P'(0)$, without loss of generality $\mathcal P(0)\ne 0$.
Thus,
\begin{equation}\label{zT}
\cosh \mu
T = z(T).
\end{equation}
The boundedness and unboundedness
of  $e^{\mu t} \mathcal P(t)$ (with its derivative and
antiderivative) is completely defined by its characteristic exponent
 $ \mu $, defined as a solution of \eqref{zT}.
 Unboundedness takes place for $|\cosh \mu T|>1$.

In particular, if $z(T)>1$, then $\mu\in \mathbb R$, and  the general solution of \eqref{P} has the form $P=C_+
e^{\mu t} \mathcal
 P(t) +C_-
e^{-\mu t} \mathcal
 P(-t)$, with arbitrary constants $C_+$, $C_-$. Thus, for an arbitrary choice of the data $P$ is unbounded and $q$ oscillates with a growing amplitude.

3. However, for a special choice of the data with $C_+=0$ the solution $P$ tends to zero as $t\to \infty$, however $q$, computed with these data, can vanish in a finite point $t_*$. Therefore, the respective solution of \eqref{uv} can blow up or not.
   If the solution does not blow up, then $u$
 and $v$ tend to zero as $t\to \infty$. If this property holds  for every $r_0\in \bar {\mathbb R}_+$, then
 $r_0 F_r(t,r_0)\to 0$, $r_0 G_r(t,r_0)\to 0$ and
  the solution of
 \eqref{4},
\eqref{CD1}
 tends to the affine one uniformly on each
compact subset of the half-axis $[0,\infty)$ as $t\to \infty$.

4. Let us show that if  $C_+=0$ for all characteristic curves, then $G(t,r)$ and $F(t,r)$ are functionally dependent, that is $\Delta(t,r)=F_r G_t-F_t G_r=0$ for every
$r, t\in \bar {\mathbb R}_+$.

Differentiating \eqref{sys_pol1} with respect to $r$ and $t$ we get
\begin{equation}
\dot \Delta =((\bd+3)F+r F_r) \Delta
\end{equation}
along the characteristic curve $r=r(t)$, starting from $r(0)=r_0$.
Therefore it is enough to check that $\Delta (0,r_0)=0$.

 Since $P(0)=p_1(0)=u_0=C_-\,\mathcal
 P(0)$, $\dot P(0)=-C_-(\dot {\mathcal P}(0)+\mu \mathcal P(0))$, $p_2(0)=v_0=C_-(\dot {\mathcal
 P}(0)+\mu \mathcal
 P(0)+\frac{\bd-2}{2}F_0)$, then
 $$
 \frac{v_0}{u_0}=\frac{G_r(0,r_0)}{F_r(0,r_0)}=\dot {\mathcal
 P}(0)+\mu \mathcal
 P(0)+\frac{\bd-2}{2}F_0.
 $$
The value of $F_0=F(0,r)$, $\mu$, $\mathcal P(0)$ also depends on $r_0$, we do not write this argument for brevity.
 We can take the values of $G_t(0,r)$ and $F_t(0,r)$ from \eqref{sys_pol1}.  After computations we get
 $$\Delta(0,r_0)=\left((1-\bd G_0) \,F_0+(F_0^2+G_0)\left(\frac{\bd-2}{2}\,F_0+\mu+\frac{\dot {\mathcal P} (0)}{\mathcal P (0)}\right)\right)F_r(0,r_0).$$
 For every $r_0$ we can choose $G_0(r_0)=G(0,r_0)$ and $F_0(r_0)=F(0,r_0)$ such that $\Delta(0,r_0)=0$, and these data correspond to a functionally dependent $G$ and $F$. In other words,  the solution corresponds to a simple wave and, in general, $F=F(G)$. This case we consider in Sec.\ref{SV}.

\bigskip

%\proof

 5. Thus, if we succeed to prove that in our case the
 characteristic exponent is real, e.g. $z(T)>1$, then we prove the theorem.

\bigskip
First we prove a particular case of Theorem \ref{MT}.

\begin{prop}\label{Pr1} The statement of Theorem \ref{MT} holds for the case of small deviation from the zero stationary
state.
\end{prop}
\proof
%Almost all non-affine small deviation from the zero
%stationary state
 %blows up in a finite time except for the only one choice
%of initial data. If a perturbation is globally smooth in time, it
%tends to the affine solution as $t\to \infty$.
Our reasoning are similar to the asymptotical analysis of the Mathieu
equation \cite {BEr67}
\begin{eqnarray*}\label{Matier}
\ddot{ p}(\tau)+(h-2 \theta\,\cos 2\tau ){
p}(\tau)=0,\quad h>0.
\end{eqnarray*}
Solution $z(\tau)$ of the Mathieu  equation with initial
conditions $z(0)=1$, $\dot z(0)=0$ is called the Mathieu C function and
can be studied asymptotically. Namely, if $|\theta|\ll 1$, then we can consider $\theta$ as a parameter of regular perturbation, and the solution
is found as a series $z=\sum\limits_{k=0}^{\infty}z_k(\tau) \theta^k $. For   every $k$ we get a linear nonhomogeneous equation $\ddot{ z_k}(\tau)+h {
z_k}(\tau)=g_k(\tau),$ subject to initial data $z_0(0)=1$, $\dot z_0(0)=0$, $z_k(0)=\dot z_k(0)=0$, $k\in \mathbb N$, which can be solved in a standard way. Evidently, $z_0(\pi)=\cos(\pi\sqrt{h})$. It is easy to check that  $z_1(\pi)=0$. Thus, we get
$$\cosh \mu \pi =z(\pi) = \cos(\pi\sqrt{h})+\sum\limits_{k=2}^{\infty}z_k(\pi) \theta^k.$$
 It is
an analog of formula
\cite {BEr67}, Sec.16.3 (2).

If $z_0(\pi)=1$, then to find the effect of a small perturbation on the characteristic exponent, we calculate $z_k(\pi)$ up to the step when $\alpha=z_k(\pi)\ne 0 $ and get
 $$\cosh \mu \pi =z(\pi) =1 +\alpha \theta^k +o(\theta^k),\,\theta\to 0, \, k\ge 2. $$
.

Our situation is more complicated, since we have to consider the periodic coefficient $Q$ in \eqref{P} as a series in
small initial deviations from zero equilibrium are
order $\varepsilon$, $\varepsilon\to 0$. Moreover, the period itself is a series in $\varepsilon$. 
We expand all functions up to the second order.
Namely, if we assume
\begin{eqnarray*}
G(t)&=& \varepsilon G_1(t)+\varepsilon^2 G_2(t)+  o(\varepsilon^2), \quad
F(t)= \varepsilon F_1(t)+\varepsilon^2 F_2(t)+  o(\varepsilon^2), \\ G(0)&=&\varepsilon,\quad F(0)=0,
\end{eqnarray*}
then system  \eqref{GF}
implies
\begin{eqnarray*}
G_1(t)& =& \cos t,\quad F_1(t) = -\sin t, \\
G_2(t)&=& -\frac12+\frac13(2+ \bd) \cos t- \frac13(2+ \bd )\cos 2 t,\\
F_2(t)&=& -\frac13(2+ \bd) \sin t+\frac13(2+ \bd)  \sin 2 t.
\end{eqnarray*}
Let us assume
\begin{eqnarray*}
z(t)= z_0+\varepsilon z_1(t)+\varepsilon^2 z_2(t)+  o(\varepsilon^2), \quad z(0)=1, \,z'(0)=0,
\end{eqnarray*}
and substitute to \eqref{P}. Then
\begin{eqnarray*}
z_0(t)& = &\cos t,\quad z_1(t) = -\frac{1}{12}(2+\bd)(2\cos(t)+\cos 2t-3), \\
z_2(t)&=& w\, t \sin t +A_0 + A_1\cos t+A_2\cos 2t+A_3\cos 3t,
\end{eqnarray*}
where $A_k=A_k(\bd)$, $k=0,\dots, 3$, $w=\frac{1}{12} (\bd-1) (\bd-4)\ne 0$.

One can check that $z_0(2\pi)=1,$ $z_1(2\pi)=z_2(2\pi)=0$.

 Lemma \ref{L4} states that the period of $Q$ is
$T=2\pi (1+w\varepsilon^2 +  o(\varepsilon^2)) $
(see \eqref{T_per}), therefore  in the order $\varepsilon^2$ we should take into account the change of period.  Since  $z'_2(2\pi)=2\pi w$, therefore $z'_2(2\pi)>0$ if $T>2\pi$ and  $z'_2(2\pi)<0$ if $T<2\pi$, as $T$ changes near $2\pi$.  Therefore we have $z_2(T)>0$, in a neighborhood of $2\pi$. A more accurate result is given by the expansion $z_2(T)=4\pi^2 w^2 \varepsilon^2 + o(\varepsilon^2)$, $\varepsilon\to 0$.
Thus, $z(T)> 1 $ for $\varepsilon\ll 1$.

The proposition is proved.
$\Box$

\bigskip

Recall that in order to prove the theorem in the general case, it is necessary to show
that $z(T)>1$, where $z(t)$ is the solution of the equation
\eqref{P} with initial conditions
   $z(0)=1$, $z'(0)=0$. Because of the complicated form of $F$ and $G$,
   we cannot hope to obtain this result analytically. Even for the much simpler Mathieu equation,
   the domains of instability
for arbitrary $h$ and $\theta$ (domains on the plane  $(h, \theta)$,
corresponding to real $\mu$) can only be found numerically
 \cite{Mathieu}.

\begin{figure}[htb!]
%\centerline{
\includegraphics[scale=0.8]{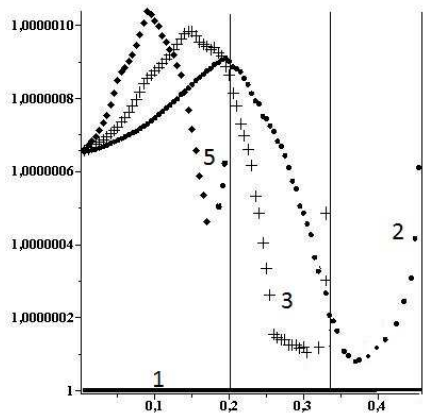}
%\vspace{-0.5 cm}
\caption{Dependence of  $e^{\mu T}$  on
$G_+$ for $\bd =1$ (and $\bd=4$ till $G_+=0.25$), solid line, $\bd=2$, solid circles, $\bd =3$, crosses,  $\bd =5$, solid diamonds. }\label{L2}
\end{figure}

However, we  can find the dependence $\mu(G_+)$,
 where $G_+$ is the greater (positive) root of the equation $F(G)=0$
 (see \eqref{fiRad2}, \eqref{fiRad3}) numerically. Namely, we need to prove that
  $e^{\mu(G_+)T(G_+)}=z(T(G_+))>1,$ for all $G_+\in (0, \frac{1}{\bd})$.

 Let us describe the steps of the computations.

 1. Given $G_+$ we find  the numerical solution of the system \eqref{GF}, $$\dot
 z=z_1,\quad\dot z_1=-Q(G,F) z,$$ with initial data $G(0)=G_+, \, F(0)= 0, \, z(0)=1, \,
 z_1(0)=0$,  the Fehlberg fourth-fifth order Runge-Kutta method was used;

 2. We find $T$ as a point where $F$ change sign, the step in $t$ is $10^{-7}$;

 3. We find $z(T)=e^{\mu T}$ (Fig.2).

Calculations are made with the step $0.005$ in $G_+$.

 Thus, we see that for $\bd\ne 1$, $\bd\ne 4$, $z(T)>1$ for all possible $G_+$, whence follows the conclusion of the theorem.

For $\bd\ne 1$ and  $\bd\ne 4$ we have  $z(T)=z(2\pi)=1$.

%We present the results of calculations only for physically
%significant cases $\bd =2$, $\bd =3$. The picture for $\bd >3$ is analogous.
 Let us notice that for large $\bd$ possible range of $G_+$ is $(0,
 \frac{1}{\bd}\ll1)$, therefore any solution can be considered as a small perturbation of the zero state.

\begin{remark}
Note that the difference $\Delta=z(T)-1$ can be regarded as a "measure of instability" in the sense that the greater this difference, the faster the solution blows up. Fig.2 shows that first, if the deviation $\varepsilon=G_+$ from the zero stationary state is small enough, $\Delta$ increases monotonically in $\varepsilon$ (from the proof of Proposition~\ref{Pr1} one can see that
$\Delta=c^2\varepsilon^4+o(\varepsilon^4)$, $c=\rm const$). However, then $\Delta$ sharply decreases, remains rather small in a narrow range of $\varepsilon$, and then increases again near the  boundary $\frac{1}{\bd}$.
  \end{remark}

 \begin{remark}
 The result about the breaking of oscillations for physical dimensions $\bd=2$ and $\bd=3$
 is confirmed by direct numerics, developed specifically for the cold plasma equations \cite{CH18}, \cite{GFCA}.
 \end{remark}

 \begin{remark}
 Numerical results, performed with high accuracy, suggest that the oscillation does not blow up for $\bd=4$ for all cases of non-negative density $G_+<\frac14$. We do not currently know of an analytical proof analogous to $\bd=1$ for this conjecture.
 This problem reduces to finding an additional first integral of the system of four equations \eqref{GF}, \eqref{lamD}.
 \end{remark}

 \section{Simple waves}\label{SV}
Theorem \ref{MT} predicts the existence of
 non-affine solutions with special initial data \eqref{CD1},
  which are globally smooth and tends to an affine solution as $t\to
  \infty$.

 To construct them, we look for simple waves of the equation
  \eqref{sys_pol1},
  such that $F=F(G)$. Thus,  \eqref{sys_pol1} reduces to one equation
\begin{eqnarray}\label{F(G)}
    \dfrac{\partial G}{\partial t}+F(G) r \dfrac{\partial G}{\partial r}=F(G)(1-\bd G),
    \end{eqnarray}
  with  $F(G)$  found from \eqref{fiRad2} or \eqref{fiRad3}, where
  the constant $C_\bd$ in \eqref{fiRad2} or \eqref{fiRad3} does not
  depend on the initial point $r_0$ and the periods of oscillations
  given as \eqref{T} are equal for all characteristics. If we fix
  $C_\bd$, we obtain the relation between $G$ and $F$ in this
  special kind of solution, and the corresponding initial data.

\begin{example}\label{Ex1} Let us construct initial data, corresponding to a
simple wave, for $\bd=2$. We choose $G_0(r)=\frac14 e^{-a^2 r^2}$.
If we complement this datum by the zero initial velocity $F_0(r)=0$,
we get a standard laser pulse \cite{CH18}. Nevertheless, these data
are not appropriate for our goal. Instead, we choose $C_2=-2$  and
find $F_0(r)$ from the condition $$ F_0^2(r)=\frac12\,((2G_0(r)-1)
\ln|1-2G_0(r)|+C_2(2G_0(r)-1) -1),$$ see \eqref{fiRad2}. It can be
readily checked that the right-hand side of this expression is
positive. Thus,
$$
F_0(r)=\pm \frac12
\,\left(1-e^{-a^2r^2}+\left(\frac12\,e^{-a^2r^2}-1\right) \,\ln
\left(1-\frac12\,e^{-a^2r^2}\right)\right)^\frac12.
$$
We can see that $F_0(r)\to\pm \frac12$ as $r\to \infty$, therefore
this initial datum can be considered as a perturbation of the affine
initial datum $\bv = \pm \frac12 \,\br$.
\end{example}

Further, differentiating \eqref{F(G)} with respect to $r$, for
$s=G_r$ we
  obtain along the characteristic curve, starting from the point $r_0$, the
  following Bernoulli  equation (we take into account \eqref{rG}):
\begin{eqnarray}\label{g}
&&\dot s \,=\,- K_2(G)\, s^2 +K_1(G)\, s, \\&& K_1(G)=F'(G)\,(1-\bd
\,G) -(1+\bd) F(G),\quad K_2=r(G)\, F'(G),\quad G=G(t),\nonumber
\end{eqnarray}
which has a trivial solution $s=0$, corresponding to the affine
solution. If $s(0)\ne 0$, then the derivative of $s(t)$ changes sign
on the graph of function $y(t)=\frac{K_1(t)}{K_2(t)}$, which is
periodic and lies from both sides of the axis $s=0$. Thus, $s(t)$
according the scenario of Theorem \ref{MT} either oscillates,
asymptotically tending to zero or blows up in a finite time.

Due to a complex structure of $G(t)$ it is not possible to obtain a
criterium of the blow-up, however, the following result holds.

We call a perturbation of a solution small if the $C^1$-norm of the
difference between the perturbation and the solution itself is small
on every compact subset of the half-axis $[0,\infty)$.

\begin{prop}\label{prop2} A sufficiently small  "simple wave" perturbation of
the affine data $(\bE_0=0, \, \bv_0=\beta \,\br) $, $\beta\ne 0 $,
generates a solution of \eqref{4}, \eqref{CD1}, which does not blow
up in a finite time.
\end{prop}

\proof The small perturbations have the form $G(t)=\epsilon \cos t
+o(\epsilon)$, $F=F(G)$, $\epsilon\to 0$. It can be checked that
this kind of perturbation is possible only if we choose
$C_2=-1-2\beta^2$ and $C_\bd =\frac{1}{\bd-2}+\beta^2$, $\bd\ge 3$,
with $\beta^2>\epsilon^2$. Note that $\beta$ is common to all points
$r_0\in[0,\infty)$, while $\epsilon$ may depend on $r_0$.

Since
$$
K_1=-(\bd+2)\,|\beta| +O(\epsilon), \quad K_2=-r_0\,|\beta|
+O(\epsilon),%\quad \epsilon\to 0,\quad \beta>0,
$$
then neglecting the terms which tends to zero with $\epsilon$ we get
from \eqref{g}
\begin{equation}\label{s}
  s(t)=\frac{(\bd +2) s(0)}{(\bd +2) e^{(\bd +2) |\beta| t}+(1- e^{(\bd
+2) |\beta|  t})\,r_0 s(0)}, \quad r_0\ge 0.
\end{equation}
Thus, we always can choose sufficiently small $|s(0)|\,r_0$ such
that the denominator of \eqref{s} does not vanish for $t>0$. Thus,
$s$ does not blow up and tends exponentially to zero. $\Box$

\begin{remark} The result, similar to Proposition \ref{prop2} can be proved
for the perturbations of an arbitrary affine data $(\bE_0=\alpha\,
\br, \, \bv_0=\beta \,\br) $, $\alpha^2+\beta^2\ne 0 $.

It also holds for perturbations in the class of simple waves of the
trivial steady state. However, in the latter case, the analysis is
more delicate. Note also that the initial perturbation of the
trivial state in the class of simple waves cannot be compactly
supported or vanish at infinity. Consequently, a solution globally
smooth in time tends as $t\to \infty$ to an affine solution, which
is itself a small perturbation of the trivial state.
\end{remark}
\begin{remark} It is not surprising that there exists a special subclass of solutions to
\eqref{4} equations that have a more regular behavior than solutions in the general case.
For one-dimensional equations of relativistic cold plasma considered in
\cite{RChZAMP21} the situation is similar, i.e. simple waves can be
globally smooth in time.
\end{remark}

\section{Behavior of density}

Since we reformulated \eqref{EP} in terms of $\bV, \bE$ as \eqref{4}, \eqref{CD1}, we now need to return to the original variables and understand how the singularity in the term of density $n$ arises.

Since $n= 1- \lambda$ and  $\lambda =\Div \bE$  in notation of Sec.\ref{Der} (see \eqref{n}), then $n\to +\infty$ as $\lambda\to -\infty$.
Recall that $\lambda=v+\bd G $, $G $ is bounded  (Lemma \ref{P2}), then $\lambda\to -\infty$ if and only if  $v\to -\infty$.

When we consider the solution along a characteristic, starting from a fixed point $r_0\ge 0$, then
according to Sec.\ref{SRadon}, $v=\frac{p_2}{q}$, where the behavior of $q$ obeys \eqref{q}, $q(0)=1$, which, in its turn, is defined by $P$. The function $P$  oscillates with exponentially increasing amplitude and in some moment $t_*$ the function $q$ becomes zero and $v$ becomes infinity. However, at $0<t<t_*$ the denominator $q$ oscillates, being positive, getting closer and closer to zero. So $n(t, r(t)) $ also oscillates before going to infinity. The behavior of $\lambda$ that determines $n$ can be studied numerically as a part of the solution of  \eqref{GF}, \eqref{lamD}.

The easiest way to trace these oscillations is  when the blow up occurs  at the point $r_0=0$, where the maximum of density forms  (the physicists call it the "axial maximum"). The respective characteristic is a straight line (see \eqref{char}). In other words, along the characteristic $r_0=0$, the solution loses smoothness earlier than along all other characteristics. For this situation the initial data should be especially chosen (as in Example \ref{Ex1}). With arbitrary initial data, the behavior of the density is complex, and a supercomputer is required for a thorough analysis. We refer to the results of computations and pictures given in the book \cite{CH18} for the physical cases $\bd=2$ and $\bd=3$, Ch.4 and 8. The density forms a delta-shape singularity.

\section*{Acknowledgments}
Supported by the Moscow Center for
Fundamental and Applied Mathematics under the agreement
¹075-15-2019-1621. The author thanks V.V.Bykov and E.V.Chizhonkov
for discussions.

\end{document}